\documentclass{elsart}

\usepackage{graphics}

\usepackage{amssymb}
\usepackage{verbatim}
\journal{Chemical Physics}
\begin{document}

\begin{frontmatter}



\title{On-line database of the spectral properties of polycyclic
aromatic hydrocarbons}

\author[label1,label2]{Giuliano Malloci\corauthref{cor}},
\corauth[cor]{Corresponding author. Tel: +39-070-675-4915;
Fax: +39-070-510171} 
\ead{gmalloci@ca.astro.it}
\author[label1]{Christine Joblin},
\author[label2]{Giacomo Mulas}

\address[label1]{Centre d'Etude Spatiale des Rayonnements \textendash{}
CNRS et Universit\'e Paul Sabatier Toulouse~3 \textendash{} Observatoire Midi-Pyr\'en\'ees,
9 Avenue du Colonel Roche, 31028 Toulouse Cedex 4, France}
\address[label2]{INAF \textendash{} Osservatorio Astronomico 
di Cagliari-Astrochemistry Group, Strada 54, Localit\'a Poggio dei Pini, 
I\textendash09012 Capoterra (CA), Italy}

\begin{abstract}
We present an on\textendash line database of computed molecular properties for
a large sample of polycyclic aromatic hydrocarbons (PAHs) 
in four charge states: -1, 0, +1, and +2. At present our database 
includes 40 molecules ranging in size from naphthalene and azulene 
(C$_{10}$H$_{8}$) up to circumovalene (C$_{66}$H$_{20}$). We performed our 
calculations 
in the framework of the density functional theory (DFT) and the 
time\textendash dependent DFT to obtain the most relevant molecular parameters 
needed for astrophysical applications. 
For each molecule in the
sample, our database presents in a uniform way the energetic,
rotational, vibrational, and electronic properties. It is freely accessible on
the web at \textsf{http://astrochemistry.ca.astro.it/database/} and
\textsf{http://www.cesr.fr/$\sim$joblin/database/}.
\end{abstract}

\begin{keyword}
 Astrochemistry \sep PAHs \sep Molecular data \sep DFT \sep TD\textendash DFT
\end{keyword}

\end{frontmatter}


\section{Introduction}
\label{introduction}

Polycyclic aromatic hydrocarbons (PAHs) are a large class of conjugated 
$\pi$\textendash electron systems of great importance in many research areas, such as 
combustion and environmental chemistry, materials science, and astrochemistry. 
In the astrophysical context, in particular, PAHs are found in carbonaceous 
meteorites \cite{hah88}, in interplanetary dust particles \cite{cle93}, and 
are tentatively identified in the atmospheres of Titan and Jupiter 
\cite{sag93} and in the coma of Halley's comet \cite{mor94cla04}. In the 
study of the interstellar medium (ISM) PAHs have been considered well before 
any IR observations of interstellar objects were available. Complex molecular 
aggregates with dimensions of $\sim$10~\AA{} were first proposed to contribute to 
the interstellar extinction \cite{pla56}. Large condensed\textendash ring hydrocarbons 
were later recognised as the ``Platt particles'', precursors of small graphite 
grains \cite{don68}.
The similarity between the vibration of C\textendash C and C\textendash H bonds in large fused\textendash ring 
aromatic molecules and on the surface of interstellar carbon particles was
then put forward to explain the origin of some emission bands observed 
in the near and mid\textendash IR in the spectra of many astronomical objects,
excited by vis\textendash UV photons \cite{dul81}. 

The observed spectra required temperatures unattainable in thermal 
equilibrium by solid\textendash state particles, suggesting molecule\textendash sized carriers, 
subject to large temperature fluctuations.
From abundance, photo\textendash stability, and spectral constraints PAHs and related 
species were thus suggested as the most natural carriers of the whole family 
of these so\textendash called ``Aromatic Infrared Bands'' (AIBs) observed near 3.3, 6.2, 
7.7, 8.6, 11.3, and 12.7~$\mu$m \cite{leg89all89}. Following the stochastic 
absorption of single vis\textendash UV photons, a typical PAH quickly converts most 
or all of the electronic excitation energy into vibrational energy of 
low\textendash lying electronic states (typically the ground\textendash state or the first excited 
state).
Due to its low heat capacity \cite{and78sel84} the molecule is transiently 
heated to peak temperatures of the order of 10$^3$~K. In a collision\textendash free 
environment it then loses its energy through the emission of radiation; IR 
fluorescence, in particular, is the proposed explanation for the observed
AIBs \cite{leg89all89}. The observations of the AROME balloon\textendash borne experiment
\cite{gia88ris94} already suggested the ubiquitous presence of AIBs in our 
Galaxy as well as in external galaxies. This is now confirmed by the large 
wealth of observations made with several space\textendash based missions, such as the 
Infrared Space Observatory (ISO)\footnote{The official ISO web page is 
\texttt{http://www.iso.vilspa.esa.es}}
and the Spitzer Space 
Telescope\footnote{\texttt{http://www.spitzer.caltech.edu}}, 
which show that the AIBs dominate the near and medium\textendash IR spectrum of many 
interstellar and circumstellar objects whose radiation field intensities 
differ by several orders of magnitude \cite{bou98pee04}. 

PAHs are estimated to account for a substantial fraction of the total 
interstellar carbon budget \cite{bou99}, and are seen as an intermediate 
stage between the gas and dust phases of the ISM, i.e., large molecules or 
very small grains. As such, they are included in some form 
in all interstellar dust models \cite{models} and are thought to play a 
role also in the chemistry \cite{chemistry}.
They are 
expected to exist in a wide variety of environments, in a complex statistical 
equilibrium of different charge and hydrogenation states \cite{baklep}.
We remark that in the astrophysical context and, consequently, in 
the astrophysical literature, the term ``PAHs'' does not correspond to the
strict definition as given in chemistry textbooks, but should be read
more like ``PAHs and related species'', including partially dehydrogenated
species and molecules in which one or more peripheral hydrogen atoms have 
been substituted by other functional groups.
Although several open questions remain opened (their formation mechanisms,
their size distribution, their charge and hydrogenation state, etc.) 
this subject has motivated a large number of observational, experimental, 
quantum\textendash chemical, and modelling studies \cite{rev}. These, in turn, led to 
the further speculation that PAHs could account for other unidentified 
spectral features of interstellar origin:
\begin{enumerate}
\item
since large neutral PAHs and all ionised PAHs absorb in the visible, they were
proposed as plausible candidate carriers of (at least) a subset of diffuse
interstellar bands (DIBs) \cite{dibs1}, about 300 absorption
features observed in the near\textendash UV, visible, and near\textendash IR spectra of reddened
stars \cite{dibs2};
\item
mixtures of gaseous PAHs (present as ions, radicals as well as neutral species)
could give an important contribution to the 217.5~nm bump and to the far\textendash UV
rise of the interstellar extinction curve \cite{job};
\item
PAHs are known to be efficient emitters in the visible through fluorescence 
and/or phosphorescence mechanisms \cite{bir70}. On the basis of matrix 
isolation spectroscopy \cite{mis} and gas\textendash phase experiments \cite{oss}, PAHs 
and their cations seem to be good candidates to account for the optical 
emission bands observed in peculiar interstellar objects \cite{rr} and 
related to some DIBs;
\item
small neutral PAHs such as anthracene
(C$_{14}$H$_{10}$) and pyrene (C$_{16}$H$_{10}$), have been tentatively identified
in the protoplanetary Red Rectangle nebula \cite{bl} and in other ordinary 
reflection nebulae \cite{vij05b}, based on the observation of a blue 
luminescence (BL), an emission band extending from 360 up to $480$~nm;
\item
PAH dications \cite{dications} are believed to contribute to the extended red 
emission (ERE), a broad emission band observed from 540 to about 900~nm in a 
large number of dusty interstellar environments \cite{ere}.
\end{enumerate}

\subsection{Collecting the spectral properties of PAHs}

The knowledge of the molecular parameters for a large sample of PAHs in all 
their relevant ionisation and hydrogenation states is of fundamental 
importance for our understanding of the physics and chemistry of the ISM 
\cite{rev}.

In the last few years we have started a long\textendash term project to produce an atlas
of synthetic spectra of \emph{individual} PAHs to be compared with 
astronomical observations both in absorption in the visible \cite{mal03} and 
in emission in the IR \cite{atlas}. To this end we need the previous knowledge 
of some key molecular parameters for a large number of species, and use them 
as a basis to run Monte\textendash Carlo simulations \cite{job02,mc} of their 
photophysics in specific interstellar environments. Such an approach enabled 
us to perform a cross\textendash check for the identification of small neutral PAHs in 
the Red Rectangle nebula \cite{bl} by comparing our quantitative prediction 
for the IR emission spectrum of each given molecule and the available ISO data 
\cite{mul06a}. In addition, these simulations provide an 
important preparatory work for the forthcoming Herschel Space Observatory 
mission\footnote{\texttt{www.sron.nl/divisions/lea/hifi/directory.html}},
in the attempt to identify specific PAHs through their low\textendash frequency 
vibrational modes involving the whole skeleton of the molecule in a 
collective vibration \cite{atlas,zha96pir06}. Resolving the rotational 
structure of these so\textendash called ``flopping'' or ``butterfly'' modes provides one 
more crucial identification element for interstellar PAHs \cite{atlas}.

Taking advantage of the great advances realised over the past years in 
computer performances, we have started a systematic theoretical study of 
the spectral properties of a large sample of PAHs in different charge states 
\cite{mal}. We used in a uniform way well\textendash known and established 
quantum\textendash chemical techniques for the study of the electronic 
ground\textendash state and excited\textendash state properties needed for astrophysical 
applications (see next Section for technical details).

Here we present a homogeneous database of the computed molecular properties 
for a sample of 40 different PAHs. For each molecule in the sample and for its 
charge states -1, 0, +1 and +2, our spectral database presents in tabular or
visual form the energetic, rotational, vibrational, and electronic properties.
This database is freely accessible on\textendash line and is suitable for further 
accumulation of new data. It can be found at the following web\textendash pages:
\begin{itemize}
\item
\textsf{http://astrochemistry.ca.astro.it/database/} \\ 
at the INAF\textendash Osservatorio Astronomico di Cagliari (Italy) and its mirror site 
at:
\item
\textsf{http://www.cesr.fr/$\sim$joblin/database/} \\ 
at the Centre d'~Etude Spatiale des Rayonnements of Toulouse (France). 
\end{itemize}
In Sect.~\ref{database} we describe the molecular properties available and the 
theoretical methods we used to obtain them (Sect.~\ref{theory}), the sample of 
different species included until present (Sect.~\ref{sample}), and the general
structure of the database (Sect.~\ref{structure}). Section~\ref{future} 
presents our concluding remarks.

\section{The database}
\label{database}

\subsection{Molecular properties and theoretical methods}
\label{theory}

Motivated mainly by basic research in astrophysics, a large number of papers 
devoted to the experimental \cite{exp1,exp2,exp3,exp4} and theoretical study
\cite{pahs1,pahs2,pahs3} of the spectral properties of PAHs and related 
species have appeared in the literature over the last decade. 
Due to the large number of electrons in the molecules considered (416 in the 
largest one, circumovalene), an \emph{ab initio} study based on the direct 
solution of the many\textendash electron Schr\"odinger equation currently has prohibitive 
computational costs. We here used the density functional theory (DFT) 
\cite{jon89} and its time\textendash dependent extension (TD\textendash DFT) \cite{mar04}, which 
are the methods of choice for the study, respectively, of the ground\textendash state 
and the excited\textendash state properties of such complex molecules as PAHs.

At present, the computed molecular properties of individual PAHs that can be 
found in our database are:
\begin{enumerate}
\item \label{item1} harmonic vibrational frequencies and integrated IR 
absorption cross\textendash sections;
\item \label{item2} rotational constants at the optimised ground\textendash state 
geometry;
\item \label{item3} electron affinities;
\item \label{item4} first and second ionisation energies;
\item \label{item5} permanent dipole moments;
\item \label{item6} vis\textendash UV photo\textendash absorption cross\textendash sections up to $\sim30$~eV.
\item \label{item7} static dipole polarisabilities;
\end{enumerate}
These are the most relevant molecular parameters needed for astrophysical 
purposes. The set of PAHs for which \emph{all} of the above data are 
simultaneously available from laboratory experiments is relatively small. 
Thanks to the ever increasing computational power available, the use of
quantum\textendash chemical tools is the next best alternative and can be furthermore 
useful as a guide for future experimental work.

All calculations for the electronic ground\textendash state have been performed using 
the Density Functional Theory module of the Gaussian\textendash based code 
\textsc{nwchem} \cite{str05}. In particular, we employed the widely used 
hybrid B3LYP functional, a combination of the Becke's three parameters 
exchange functional \cite{bec93} and the Lee\textendash Yang\textendash Parr gradient\textendash corrected 
correlation functional \cite{lee88}. We first optimised the ground\textendash state 
geometries using the relatively inexpensive \mbox{4\textendash31G} basis set, followed 
by the full vibrational analyses (item~\ref{item1}) to confirm the geometries 
obtained to be global minima on the potential energy surface. The combination 
\mbox{B3LYP/4\textendash31G} was shown to give good results in the study of the 
vibrational properties of PAHs, scaling all frequencies by the same empirical 
scale\textendash factor \cite{lan96,bau97}. We thus adopted the same uniform scaling
procedure derived by these authors for all of the PAHs under study in their 
-1, 0, +1 and +2 charge states. Since for  open\textendash shell systems analytical 
second derivatives are not yet implemented in the version of \textsc{nwchem} 
that we used (version 4.7), we performed the vibrational analyses for 
all the anions, cations, and dications triplets, using the \textsc{Gaussian03} 
quantum chemistry package \cite{g03}.


We then started from the optimal geometries and the corresponding 
self consistent field solution obtained in the previous step to refine the 
optimisation with the larger \mbox{6\textendash31+G$^\star$} basis\textendash set, a valence double 
zeta basis set augmented with $d$ polarisation functions and $s$ and $p$ 
diffuse functions to each carbon atom \cite{fri84}. The rotational constants 
(item~\ref{item2}) have been derived from the optimised 
B3LYP/\mbox{6\textendash31+G$^\star$} geometries.
Although basis set convergence is not yet expected at this level, in view of 
the large systems under study some compromise between accuracy and 
computational costs had to be made.

We thus computed the adiabatic and vertical values of both the electron 
affinity (item \ref{item3}) and the first and second ionisation energy 
(item \ref{item4}). Adiabatic values are evaluated as the difference between 
the total energies of the neutral and the corresponding ion minima, 
respectively; these quantities, therefore, take into account the structural 
relaxation of the molecule following the ionisation process. The vertical 
values, computed at the optimised geometry of the neutral molecule, neglect 
structural relaxation but include wavefunctions relaxation due to the addition 
of one electron  (anion) or the removal of one (cation) or two (dication)
electrons.
All neutral and singly\textendash ionised species were computed as singlet and doublet,
respectively, while for dications we computed both their singlet and triplet
ground states. This enabled us to predict the relative energies of electronic
states having different multiplicities. For almost all of the molecules under
study (34 out of a total of 40), our calculations predict the dication overall
ground\textendash state to be the singlet \cite{mal06}. 
Whenever available, we list the experimental electron affinities and single 
ionisation energies as reported in the NIST Chemistry WebBook \cite{lia05}. 
As to the second ionisation energies we report the values obtained in the only 
available experimental measurements of the second ionization energies of the 
PAHs we considered \cite{tob94,sch01,den06}. For polar species we list also 
the corresponding permanent dipole moments (item \ref{item5}), which might be 
relevant for the astronomical search of PAHs in the microwave region through 
their pure rotational spectrum \cite{lov05}. 

Concerning item~\ref{item6} in the above list, we already presented a 
collection of individual vis\textendash UV photo\textendash absorption spectra computed in a 
uniform way for a large sample of anionic, neutral, cationic, and dicationic 
PAHs \cite{mal}.  We used a real\textendash time real\textendash space implementation of the 
Time\textendash Dependent Density Functional Theory (TD\textendash DFT) \cite{yab99} as given in 
the \textsc{octopus} computer code \cite{mar03}. 
Our results are in good agreement with the only available measurements of 
photo\textendash absorption cross\textendash sections of neutral PAHs from the visible to 
the vacuum\textendash UV spectral domain \cite{job}. While experimental spectra in the 
far\textendash UV are needed for a similar direct validation for PAH ions, our 
theoretical data for neutral species reproduce the overall far\textendash UV behaviour, 
including the broad absorption peak dominated by $\sigma^*\gets\sigma$ transitions, which 
matches well both in position and width. The computed vertical excitations of 
$\pi\to\pi^\star$ character falling in the low\textendash energy range are expected to be precise 
to within a few tenths of an eV, which are the typical accuracies of TD\textendash DFT 
calculations using the currently available exchange\textendash correlation functionals 
\cite{hir99,hir03}. With respect to other theoretical studies of the 
electronic excitations of PAHs, the main step forward achieved by our work is 
represented by the spectral range covered, that extends up to $\sim$30~eV. This 
piece of information has proven to be particularly useful for astrophysical 
modelling when laboratory data are missing \cite{mul06a,mul06b,mul06c}. On the 
other hand, the main drawback is that we do not obtain independent information 
for each excited electronic state, such as its symmetry and its description in 
terms of promotion of electrons in a one\textendash electron picture. With the 
\textsc{octopus} code we also obtain the static dipole polarisabilities 
(item \ref{item7}), whose knowledge is important in the study of chemical 
reactivity. 

More technical details about the calculations 
we performed with both the \textsc{octopus} and \textsc{nwchem} codes, as well 
as on the expected accuracy achieved, can be found elsewhere \cite{mal}.



\subsection{The sample of molecules}
\label{sample}

A wide range of PAHs and related species are thought to exist in the ISM 
\cite{tie05}. The total of 40 molecules selected in this work, ranging in size 
from naphthalene and azulene (C$_{10}$H$_{8}$) to circumovalene (C$_{66}$H$_{20}$), 
cover an ample range of structures inside the two large classes of 
pericondensed and catacondensed species. Although non\textendash compact PAHs are 
expected to be less stable than compact ones with the same number of benzenoid 
rings, we extended the present study to both classes, including also large 
oligoacenes such as pentacene (C$_{22}$H$_{14}$) and hexacene (C$_{26}$H$_{16}$). 
In addition we included some PAHs with five\textendash membered rings, namely fluorene 
(C$_{13}$H$_{10}$), fluoranthene (C$_{16}$H$_{10}$) and corannulene (C$_{20}$H$_{10}$). 
Whenever possible, the known symmetry of the molecular geometry was 
assumed during calculations, e.g.,
C$_{2v}$ for phenanthrene (C$_{14}$H$_{10}$), C$_{2h}$ for chrysene (C$_{18}$H$_{12}$), 
D$_{3h}$ for triphenylene (C$_{18}$H$_{12}$), D$_{2h}$ for ovalene (C$_{32}$H$_{14}$), 
D$_{6h}$ for coronene (C$_{24}$H$_{12}$), hexabenzocoronene (C$_{42}$H$_{18}$) and 
circumcoronene (C$_{54}$H$_{18}$) and so forth). 
In the case of some of the molecules with a more than 2\textendash fold
symmetry axis, symmetry breaking was observed upon single and double
ionization, as expected from Jahn\textendash Teller distortion \cite{kat99}. 
More specifically, we obtained a symmetry reduction from D$_{6h}$ to D$_{2h}$ 
point group for coronene, hexabenzoronene, and circumcoronene, from  D$_{3h}$ 
to C$_{2v}$ for triphenylene and from C$_{5v}$ to  C$_{s}$ for corannulene.

Although 
some studies show interstellar PAHs to be on average larger than the species 
we considered \cite{bou99}, we restricted ourselves to molecules containing 
up to a maximum of 66~carbon atoms, since computational costs steeply scale 
with dimensions (e.g., about 3000 CPU hours on an IBM SP5 for the 
circumovalene ions, which were the most expensive species considered).

\subsection{General structure of the database}
\label{structure}

The main\textendash pages \textsf{http://astrochemistry.ca.astro.it/database/} and \\
\textsf{http://www.cesr.fr/$\sim$joblin/database/} contain the full list of 
molecules considered; they are ordered by incrasing masses. Direct links 
to the official web\textendash pages of the packages and computing resources used are 
given.
A separate page can then be accessed for each given species. The latter 
includes a graphical representation of the selected molecule and the links to 
the following pages:
\begin{enumerate}
\item[(i)] 
The corresponding NIST web\textendash page (if available).
\item[(ii)]
The html table of the harmonic vibrational analyses for each of the four 
charge states considered (item \ref{item1} in the list of the previous 
section); a pdf version of the table can also be downloaded. 
\item[(iii)] 
The table of the general parameters of the molecule (items \ref{item2} to 
\ref{item6}) with the addition of its symmetry properties 
(symmetry properties, irreducible representations $\Gamma_\mathrm{3N}$ and 
$\Gamma_\mathrm{vib}$, and IR\textendash active modes).
\item[(iv)]
The plots of the vis\textendash UV photo\textendash absorption cross\textendash sections (item \ref{item7})
for each charge\textendash state; all of the spectra can also be downloaded in ascii 
format.
\item[(v)] 
A list with some of the more representative papers related to the given 
molecule; more than 500 references are currently included, each of 
them pointing to the official web\textendash page of the journal (when available). 
 
\end{enumerate}

As to the last point, in particular, many references are likely to have 
been missed in our bibliographic search. We will be thankful to anyone who 
will send us more relevant references.

\subsection{Related links}

There are several other internet resources reporting relevant data and 
bibliographic references on the spectral properties of PAHs. Among them, we 
refer the reader to the official web-page of the Astrochemistry Laboratory 
at NASA Ames Research Center \textsf{http://www.astrochemistry.org/} 
and the Polycyclic Aromatic 
Hydrocarbon Structure Index \textsf{http://ois.nist.gov/pah/} and the 
Chemistry WebBook \textsf{http://webbook.nist.gov/chemistry/}
of the National Institute of Standards and Technology. Up to date references
and direct links to personal web pages can also be found in the web-page
\textsf{http://www.astrochymist.org/}, which is devoted to the field of 
astrochemistry in general.

\section{Future perspectives}
\label{future}

In the near future we will extend our database towards two main directions.
We will include larger PAHs and, at the same time, we will study some
derivatives of the molecules already considered. In particular, we plan to 
study de\textendash hydrogenated \cite{vuo00,dul06} or super\textendash hydrogenated species 
\cite{ber96sno98}, and species including heteroatoms such as oxygen 
\cite{bau98b}, nitrogen \cite{polar} or organometallic compounds
\cite{szc06,ell99}. 
Better levels of theory (e.g., larger bases and/or newly developed 
exchange\textendash correlation functionals) will be employed for the calculation of some
of the molecular 
properties considered. In addition, we plan to compute other poorly known 
molecular parameters, such as the anharmonic correction to vibrational terms 
and rotational constants \cite{bar} and the vibronic structure of electronic 
absorption and emission spectra \cite{die}. On a parallel track, all of 
these data will be used to produce synthetic spectra to compare with actual 
astronomical observations. Of course, external contributions to be included 
in the database, either experimental or theoretical results, are also very 
welcome. We therefore invite anybody interested in having their results 
made available on our database to submit their contributions following the 
contact links in the on\textendash line web\textendash pages; the appropriate 
citation to the owner of the data will be reported.

\section*{Acknowledgements}
G.~Malloci aknowledges the financial support by INAF\textendash Ossevatorio Astronomico 
di Cagliari and the ``Minist\`ere de la Recherche''. We warmly thank the authors 
of \textsc{octopus} for making their code available under a free license. We 
acknowledge the High Performance Computational Chemistry Group for the use of 
\textsc{nwchem}, 
A Computational Chemistry Package for Parallel Computers, Version~4.7 
(2005), PNNL,  Richland, Washington, USA. Part of the simulations were 
carried out at CINECA (Bologna) and CALMIP (Toulouse) 
supercomputing facilities.








\end{document}